\documentclass[prc,twocolumn,superscriptaddress,showpacs,twoside,floatfix]
{revtex4-1}

\usepackage{amssymb,epsfig}
\usepackage{color}
\usepackage{mathtools}
\begin{document}

\title{Towards the Limits of Existence of Nuclear Structure:
Observation and First spectroscopy of the Isotope $^{31}$K by measuring its three-proton Decay}

\author{D.~Kostyleva}
\email{Corresponding author: D.Kostyleva@gsi.de}
\affiliation{GSI Helmholtzzentrum  f\"{u}r Schwerionenforschung GmbH, 64291
Darmstadt, Germany}
\affiliation{II.Physikalisches Institut, Justus-Liebig-Universit\"at, 35392 Gie{\ss}en, Germany}

\author{I.~Mukha}
%\email{I.Mukha@gsi.de}
\affiliation{GSI Helmholtzzentrum  f\"{u}r Schwerionenforschung GmbH, 64291
Darmstadt, Germany}

\author{L.~Acosta}
\affiliation{INFN, Laboratori Nazionali del Sud,
Via S.~Sof\'ia, 95123 Catania, Italy}
\affiliation{Instituto de F\'isica, Universidad Nacional Aut\'onoma de M\'exico, M\'exico, Distrito Federal 01000, Mexico}

\author{E.~Casarejos}
\affiliation{University of  Vigo, 36310 Vigo, Spain}

\author{V.~Chudoba}
\affiliation{Flerov Laboratory of Nuclear Reactions, JINR,  141980 Dubna, Russia}
\affiliation{Institute of Physics, Silesian University Opava, 74601 Opava, Czech Republic}

\author{A.A.~Ciemny}
\affiliation{Faculty of Physics, University of Warsaw, 02-093 Warszawa, Poland}

\author{W.~Dominik}
\affiliation{Faculty of Physics, University of Warsaw, 02-093 Warszawa, Poland}

\author{J.A.~Due\~nas}
\affiliation{Departamento de Ingenieria Electrica y Centro de Estudios Avanzados en Fisica, Matem\'{a}ticas y Computaci\'{o}n, Universidad de Huelva, 21071 Huelva, Spain}

\author{V.~Dunin}
\affiliation{Veksler and Baldin Laboratory of High Energy Physics, JINR, 141980 Dubna,
Russia}

\author{J.~M.~Espino}
\affiliation{Department of Atomic, Molecular and Nuclear Physics, University of Seville, 41012 Seville, Spain}

\author{A.~Estrad\'{e}}
\affiliation{University of Edinburgh, EH1 1HT Edinburgh, United Kingdom}

\author{F.~Farinon}
\affiliation{GSI Helmholtzzentrum  f\"{u}r Schwerionenforschung GmbH, 64291
Darmstadt, Germany}

\author{A.~Fomichev}
\affiliation{Flerov Laboratory of Nuclear Reactions, JINR,  141980 Dubna, Russia}

\author{H.~Geissel}
\affiliation{GSI Helmholtzzentrum  f\"{u}r Schwerionenforschung GmbH, 64291
Darmstadt, Germany}
\affiliation{II.Physikalisches Institut, Justus-Liebig-Universit\"at, 35392 Gie{\ss}en, Germany}

\author{A.~Gorshkov}
\affiliation{Flerov Laboratory of Nuclear Reactions, JINR,  141980 Dubna, Russia}

\author{L.V.~Grigorenko}
\affiliation{Flerov Laboratory of Nuclear Reactions, JINR,  141980 Dubna, Russia}
\affiliation{National Research Nuclear University ``MEPhI'',
115409 Moscow, Russia}
\affiliation{National Research Centre ``Kurchatov Institute'', Kurchatov
square 1, 123182 Moscow, Russia}

\author{Z.~Janas}
 \affiliation{Faculty of Physics,  University of Warsaw, 02-093 Warszawa,
 Poland}

\author{G.~Kami\'{n}ski}
\affiliation{Heavy Ion Laboratory, University of Warsaw, 02-093 Warszawa,
Poland}
\affiliation{Flerov Laboratory of Nuclear Reactions, JINR,  141980 Dubna,
Russia}

\author{O.~Kiselev}
\affiliation{GSI Helmholtzzentrum  f\"{u}r Schwerionenforschung GmbH, 64291
Darmstadt, Germany}

\author{R.~Kn\"{o}bel}
\affiliation{GSI Helmholtzzentrum  f\"{u}r Schwerionenforschung GmbH, 64291
Darmstadt, Germany}
\affiliation{II.Physikalisches Institut, Justus-Liebig-Universit\"at, 35392 Gie{\ss}en, Germany}

\author{S.~Krupko}
\affiliation{Flerov Laboratory of Nuclear Reactions, JINR,  141980 Dubna, Russia}

\author{M.~Kuich}
\affiliation{Faculty of Physics, Warsaw University of Technology, 00-662 Warszawa, Poland}
 \affiliation{Faculty of Physics,  University of Warsaw, 02-093 Warszawa,
 Poland}

\author{Yu.A.~Litvinov}
\affiliation{GSI Helmholtzzentrum  f\"{u}r Schwerionenforschung GmbH, 64291
Darmstadt, Germany}

\author{G.~Marquinez-Dur\'{a}n}
\affiliation{Department of Applied Physics, University of Huelva, 21071 Huelva, Spain}

\author{I.~Martel}
\affiliation{Department of Applied Physics, University of Huelva, 21071 Huelva, Spain}

\author{C.~Mazzocchi}
\affiliation{Faculty of Physics, University of Warsaw, 02-093 Warszawa, Poland}

\author{C.~Nociforo}
\affiliation{GSI Helmholtzzentrum  f\"{u}r Schwerionenforschung GmbH, 64291
Darmstadt, Germany}

\author{A.~K.~Ord\'{u}z}
\affiliation{Department of Applied Physics, University of Huelva, 21071 Huelva, Spain}

\author{M.~Pf\"{u}tzner}
\affiliation{Faculty of Physics, University of Warsaw, 02-093 Warszawa, Poland}
\affiliation{GSI Helmholtzzentrum  f\"{u}r Schwerionenforschung GmbH, 64291
Darmstadt, Germany}

\author{S.~Pietri}
\affiliation{GSI Helmholtzzentrum  f\"{u}r Schwerionenforschung GmbH, 64291
Darmstadt, Germany}

\author{M.~Pomorski}
 \affiliation{Faculty of Physics,  University of Warsaw, 02-093 Warszawa,
 Poland}

\author{A.~Prochazka}
\affiliation{GSI Helmholtzzentrum  f\"{u}r Schwerionenforschung GmbH, 64291
Darmstadt, Germany}

\author{S.~Rymzhanova}
\affiliation{Flerov Laboratory of Nuclear Reactions, JINR,  141980 Dubna, Russia}

\author{A.M.~S\'{a}nchez-Ben\'{i}tez}
\affiliation{Centro de Estudios Avanzados en F\'{i}sica, Matem\'{a}ticas y
Computaci\'{o}n (CEAFMC), Department of Integrated Sciences, University of
Huelva,  21071 Huelva, Spain}

\author{C.~Scheidenberger}
\affiliation{GSI Helmholtzzentrum  f\"{u}r Schwerionenforschung GmbH, 64291
Darmstadt, Germany}
\affiliation{II.Physikalisches Institut, Justus-Liebig-Universit\"at, 35392 Gie{\ss}en, Germany}

%\author{P.~Sharov}
%\affiliation{Flerov Laboratory of Nuclear Reactions, JINR,  141980 Dubna, Russia}

\author{H.~Simon}
\affiliation{GSI Helmholtzzentrum  f\"{u}r Schwerionenforschung GmbH, 64291
Darmstadt, Germany}

\author{B.~Sitar}
\affiliation{Faculty of Mathematics and Physics, Comenius University, 84248 Bratislava,
Slovakia}

\author{R.~Slepnev}
\affiliation{Flerov Laboratory of Nuclear Reactions, JINR,  141980 Dubna, Russia}

\author{M.~Stanoiu}
\affiliation{IFIN-HH, Post Office Box MG-6, Bucharest, Romania}

\author{P.~Strmen}
\affiliation{Faculty of Mathematics and Physics, Comenius University, 84248 Bratislava,
Slovakia}

\author{I.~Szarka}
\affiliation{Faculty of Mathematics and Physics, Comenius University, 84248 Bratislava,
Slovakia}

\author{M.~Takechi}
\affiliation{GSI Helmholtzzentrum  f\"{u}r Schwerionenforschung GmbH, 64291
Darmstadt, Germany}

\author{Y.K.~Tanaka}
\affiliation{GSI Helmholtzzentrum  f\"{u}r Schwerionenforschung GmbH, 64291
Darmstadt, Germany}
\affiliation{University of Tokyo, 113-0033 Tokyo, Japan}

\author{H.~Weick}
\affiliation{GSI Helmholtzzentrum  f\"{u}r Schwerionenforschung GmbH, 64291
Darmstadt, Germany}

\author{M.~Winkler}
\affiliation{GSI Helmholtzzentrum  f\"{u}r Schwerionenforschung GmbH, 64291
Darmstadt, Germany}

\author{J.S.~Winfield}
\affiliation{GSI Helmholtzzentrum  f\"{u}r Schwerionenforschung GmbH, 64291
Darmstadt, Germany}

\author{X.~Xu}
\affiliation{Institute of Modern Physics, Chinese Academy of Sciences, Lanzhou 730000, China}
\affiliation{II.Physikalisches Institut, Justus-Liebig-Universit\"at, 35392 Gie{\ss}en, Germany}
\affiliation{GSI Helmholtzzentrum  f\"{u}r Schwerionenforschung GmbH, 64291
Darmstadt, Germany}

\author{M.V.~Zhukov}
\affiliation{Department of Physics, Chalmers University of Technology, S-41296 G\"oteborg, Sweden}

\date{\today. {\tt File: K31-draft.tex }}

\begin{abstract}
The most-remote from stability isotope $^{31}$K, which is located four atomic mass units beyond the proton drip line, has been observed. It is  unbound in respect to  three-proton (\emph{3p}) emission, and its decays  have been detected in flight by measuring trajectories of all decay  products  using  micro-strip detectors. The  $3p$-emission processes have been studied by means of angular correlations  $^{28}$S+3$p$ and the respective decay vertexes. The energies of the  previously-unknown ground and excited states of $^{31}$K  have been determined. This provides its $3p$ separation-energy value $S_{3p}$ of $-4.6(2)$ MeV.
Upper half-life limits of 10 ps of the observed $^{31}$K states have been derived from
distributions of the measured decay vertexes.

\end{abstract}


\maketitle

%===============================================================================

%\section{Introduction}

%===============================================================================
%\emph{Motivation.}
 In recent experimental \cite{Mukha:2018} and theoretical \cite{Grigorenko:2018} studies of the lightest isotopes in the argon and chlorine isotope chains, limits of existence of the corresponding nuclear structure were addressed.
  For the issue of existence of nuclear structure, we adopt the approach used in Ref.\ \cite{Grigorenko:2018}. Namely, a nuclear configuration has an individual structure with at least one distinctive state, if the orbiting valence protons of the system are reflected from the corresponding nuclear barrier at least one time. Thus the nuclear half-life may be used as a criterion here, and two extreme cases can be mentioned. The very long-lived particle-emitting states may be considered as \emph{quasistationary}. For example, the half-lives of all known heavy two-proton ($2p$) radioactivity precursors (\emph{e.g.}, $^{45}$Fe, $^{48}$Ni, $^{54}$Zn, $^{67}$Kr) are a few milliseconds \cite{Pfutzner:2002,Pomorski:2011,Blank:2005,Goigoux:2016}. For such long-lived states,   modifications of nuclear structure by coupling with  continuum  are negligible. In the opposite case of  very short-lived unbound ground states (g.s.), the continuum coupling becomes increasingly important, which can be regarded as a transition to continuum dynamics. For example, the discussion of the tetra-neutron (\emph{4n}) system  has shown that its spectrum is strongly affected both by the reaction mechanism and by the initial nuclear
structure of the participants \cite{Grigorenko:2004}.
In Ref.~\cite{Grigorenko:2018}, the isotopes $^{26}$Ar and $^{25}$Cl were predicted as the most remote nuclear configurations with identified  g.s.. Similar predictions allow to expect a number of previously-unknown unbound isotopes located within a relatively broad (by 2--5 atomic mass units) area along the proton drip line. For more exotic nuclear systems beyond such a domain, no g.s.\ of isotopes (and thus no new isotope identification) are expected. Therefore a new borderline indicating the limits of existence of isotopes in the nuclear chart and the transition to chaotic-nucleon matter may be inspected.

In this work we continue the ``excursion beyond the proton dripline'' of Ref.\ \cite{Mukha:2018} by presenting the results of additional analysis of the data obtained with a $^{31}$Ar secondary beam  \cite{Xu:2018}.
In the experiment, described in detail in Refs.\ \cite{Xu:2018, Mukha:2018}, the $^{31}$Ar beam was produced by the fragmentation of a primary 885 MeV/u $^{36}$Ar beam at the SIS-FRS facility at GSI (Germany). The previous objectives of the experiment were studies of $2p$ decays of $^{29,30,31}$Ar isotopes. We briefly repeat the general description of the experiment and the detector performance.
The FRS was operated with an ion-optical settings in a separator-spectrometer mode, where the first half of the FRS was set for separation and focusing of the radioactive beams on a secondary target in the middle of the FRS, and the second half of FRS was set for the detection of heavy-ion decay products. The secondary 620 MeV/u $^{31}$Ar beam with an intensity of 50 ions $\text{s}^{-1}$  bombarded a 27-mm thick $^9$Be secondary target located at the FRS middle focal plane.  In the cases addressed in Refs.\ \cite{Xu:2018,Mukha:2018}, the $^{29,30}$Ar nuclei were produced via neutron knockout reactions from the $^{31}$Ar ions. The decay products of unbound $^{29,30}$Ar nuclei were tracked by a double-sided silicon micro-strip detector (DSSD) array placed just downstream of the secondary target.
Four large-area DSSDs \cite{Stanoiu:2008} were employed to measure hit coordinates of the protons and the recoil heavy ions (HI), resulting from the in-flight decays of the studied $2p$ precursors. The high-precision position measurement by DSSDs allowed for reconstruction of all fragment trajectories, which let us to derive the decay vertex together with  angular HI-$p$ and HI-$p_1$-$p_2$ correlations. For example, the trajectories of measured $^{28}$S+$p$+$p$ coincidences served for the analysis, and the  spectroscopic information on $^{30}$Ar was concluded \cite{Xu:2018}. The spectra of $^{30}$Ar
were observed by using \emph{2p} angular correlations as function of their root-mean-square angle relative to $^{28}$S,
  \begin{equation}\label{eq:1}
\rho_\theta = \sqrt{\theta^2_{p1-^{28}S}+\theta^2_{p2-^{28}S}}.
 \end{equation}
%$\rho_\theta$=$\sqrt{\theta^2_{p1-^{28}S}+\theta^2_{p2-^{28}S}}$.
%
A number of by-product results was obtained in a similar way from the data recorded in the same experiment. In particular,  a \emph{3p}-unbound nuclear system of $^{31}$K was populated in a charge-exchange reaction. This mechanism has lower cross section than knockout reactions, and the obtained data have smaller statistics than in the previously-mentioned case \cite{Xu:2018}.  In spite of poor statistics with few events registered, we have obtained several nuclear-structure conclusions from the data.
The $^{31}$K spectrum was derived from trajectories of all decay products $^{28}$S+$p_1$+$p_2$+$p_3$ measured in four-fold coincidence.
All detector calibrations  were taken from the analyses reported in Refs.\ \cite{Xu:2018,Mukha:2018}.

%===============================================================================

%-------------------------------------------------------------------------------
\begin{figure}[t]
\begin{center}
\includegraphics[width=0.45\textwidth]{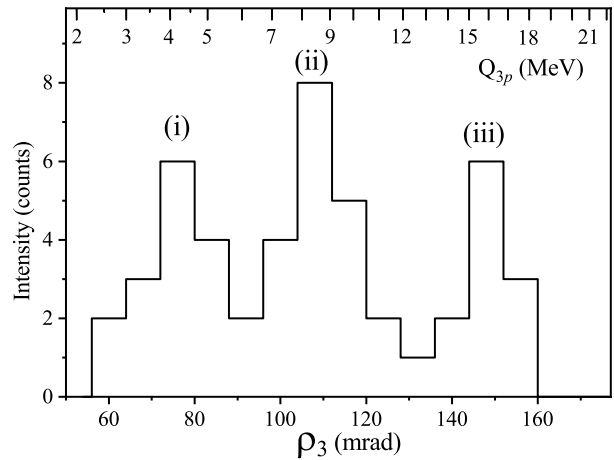}
\end{center}
\caption{ Three-proton angular correlations as function of their root-mean-square angle
$\rho_3$ [see eq.~(2)]
%$\rho_3$=$\sqrt{\smashoperator[r]{\sum_{i=1}^3} \theta^2_{p_i-^{28}S}}$
derived from the measured trajectories of all decay products,
 $^{28}$S+$3p$ (histogram), which reflect the excitation spectrum of the isotope $^{31}$K.  The peaks (i), (ii), (iii) suggest the  $^{31}$K states  whose  $3p$-decay energies $Q_{3p}$ are shown in the
upper axis.}
 \label{fig:31K-rho3}
\end{figure}
%-------------------------------------------------------------------------------

  In Fig.~\ref{fig:31K-rho3}, we present \emph{3p} correlations observed in decays of $^{31}$K as function of their root-mean-square angle
  \begin{equation}
    \rho_3 = \sqrt{\theta^2_{p1-^{28}S}+\theta^2_{p2-^{28}S}+\theta^2_{p3-^{28}S}}
\label{eq:2}
  \end{equation}
 derived from the measured  trajectories of  $^{28}$S+$3p$ coincident events. The kinematical variable $\rho_3$ is introduced because the decay protons share the \emph{3p}-decay energy, in analogy with \emph{2p} decays [see eq.~(1)]. One can see three peaks (i), (ii) and (iii) reflecting the population of  states in  $^{31}$K isotope,  and their respective $3p$-decay energies $Q_{3p}$ of about 4.5, 9 and 16 MeV may be estimated from the upper axis.
%By using these areas as selection gates, one may produce respective angular $\theta_{p-^{28}S}$ correlations projected
% from the measured  $^{28}$S+$3p$ coincidences, which is shown in Fig.~\ref{fig:theta-p-S}.

%-------------------------------------------------------------------------------
\begin{figure}[h!]
\begin{center}
\includegraphics[width=0.48\textwidth]{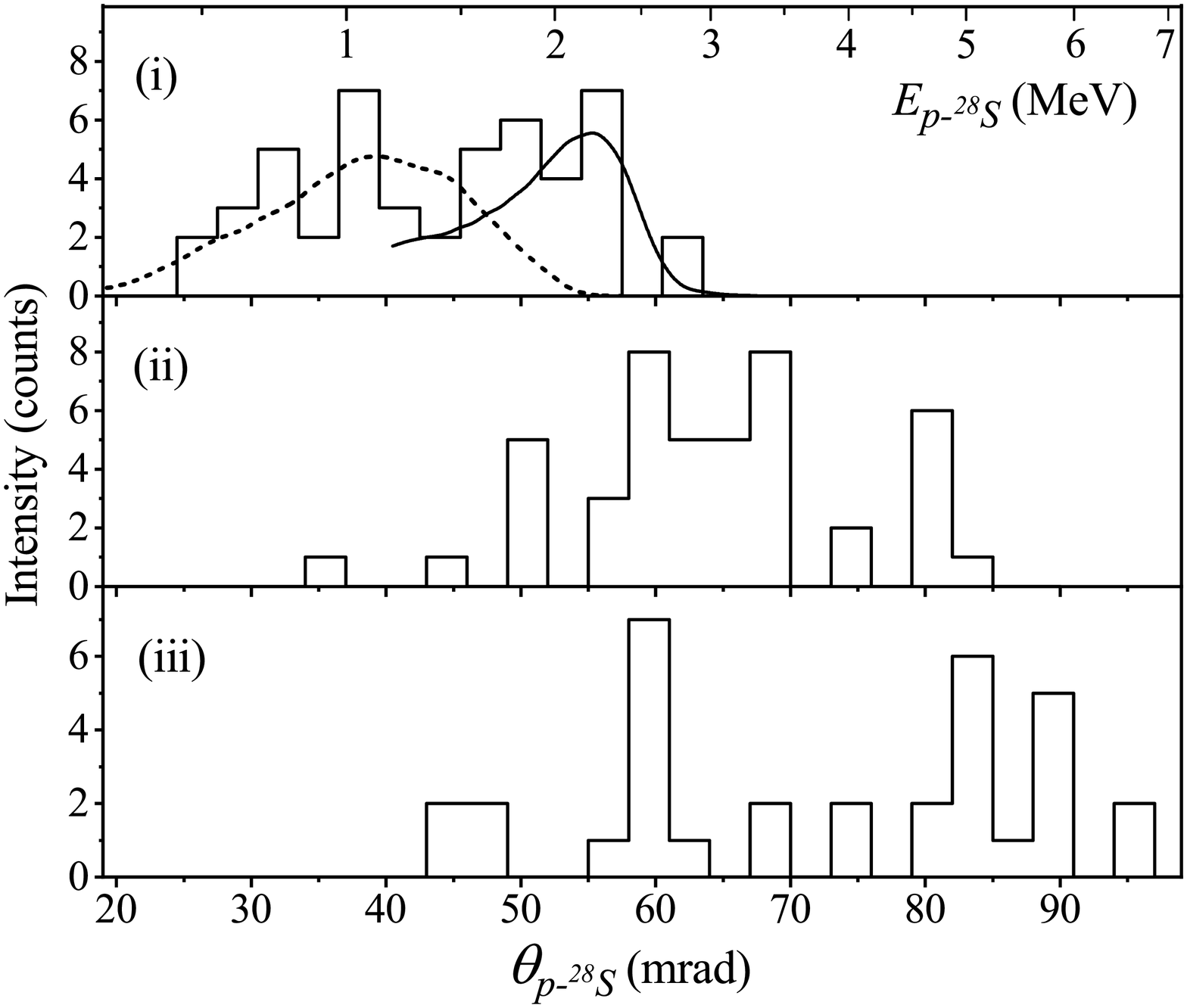}
\end{center}
 \caption{ Angular $\theta_{p-^{28}S}$ correlations projected from the measured
$^{28}$S+$3p$ coincidences (histograms). The data in panels (i), (ii), (iii) are selected by the  respective gates around the $\rho_3$-peaks in Fig.~\ref{fig:31K-rho3}.  The corresponding $1p$-decay energies $E_{p-^{28}S}$ are
given by the upper axis. The solid curve is the best-fit contribution from the initial \emph{1p}-decay of $^{31}$K into the $^{30}$Ar g.s.\ with the fitted decay energy of 2.15(15) MeV. The contribution of a subsequent \emph{2p}-decay of $^{30}$Ar with the known energy of 2.45 MeV \cite{Xu:2018} is shown by the dotted curve.
}
 \label{fig:theta-p-S}
\end{figure}
%-------------------------------------------------------------------------------

In order to establish decay schemes of the  states in $^{31}$K, we have produced angular $\theta_{p-^{28}S}$ correlations
projected from the $^{28}$S+$3p$ events which are selected by the gates around the $\rho_3$-peaks (i), (ii), (iii) in Fig.~\ref{fig:31K-rho3}. These projections are shown in the respective panels in Fig.\ \ref{fig:theta-p-S}.
In particular, the lowest-energy 4.5 MeV peak (i) may correspond to the $^{31}$K g.s., which decays by emission of a proton first into an intermediate $^{30}$Ar g.s., whose \emph{2p}-decay energy of 2.45(15) MeV is known \cite{Xu:2018}. Then the corresponding $\theta_{p-^{28}S}$ correlations in Fig.~\ref{fig:theta-p-S}(i) should consist of two contributions. The firstly-emitted proton should cause a peak in the observed $\theta_{p-^{28}S}$ correlations. The second component should be the known broad
$\theta_{p-^{28}S}$ distribution from the $^{30}$Ar g.s.\ \emph{2p}-decay \cite{Xu:2018}, which is centered at  $E_{p-^{28}S}\simeq$1.2 MeV (because the 2 protons, which can not be distinguished, share the \emph{2p}-decay energy).
We have fitted the data by a sum of two respective components:  1) the Monte-Carlo simulation including the  response of the experimental setup to \emph{1p}-emission by $^{31}$K (the simulation procedure is described in details in Refs.~\cite{Mukha:2007,Agostinelli:2003}; 2) the known detector response to the \emph{2p}-decay of $^{30}$Ar g.s.\ (see Ref.~\cite{Xu:2018}). One may see, that the small-angle region of the $\theta_{p-^{28}S}$ distribution agrees with the \emph{2p}-decay of $^{30}$Ar g.s.\ (the dotted-line taken into account the literature value of the \emph{2p}-decay energy of 2.45$^{+0.05} \!\!\!\!\!\!\!\!\!\!\!\!_{-0.10}$~MeV), while the large-angle correlations can be described by the \emph{1p}-emission of $^{31}$K into $^{30}$Ar g.s.\ (solid line) with the best-fit decay energy of 2.15(15) MeV.
%
%-------------------------------------------------------------------------------
\begin{figure}[h!]
\begin{center}
\includegraphics[width=0.3\textwidth]{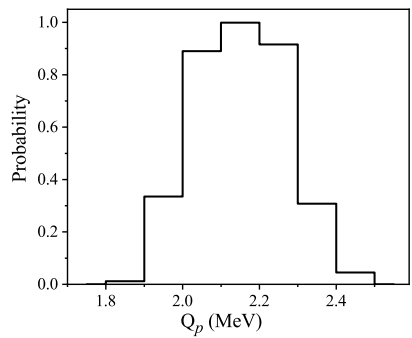}
\end{center}
 \caption{ Probability that the simulated response of the setup to the \emph{1p} decay of $^{31}$K into the $^{30}$Ar g.s.\ matches the measured angular $\theta_{p-^{28}S}$ correlations [shown in Fig.~\ref{fig:theta-p-S}(i)] as a function of assumed \emph{1p}-decay energy $Q_p$.
}
 \label{fig:K31gs_prob}
\end{figure}
%-------------------------------------------------------------------------------
%
The illustration of  procedure of the data fit is given in Fig.~\ref{fig:K31gs_prob}, where the 
probability that the simulated response of the setup to the \emph{1p} decay of $^{31}$K into the $^{30}$Ar g.s.\ matches the measured angular $\theta_{p-^{28}S}$ correlations is shown in dependence on the \emph{1p}-decay energy. The best-fit energy and its uncertainty have been derived from the distribution centroid and width, respectively.
Thus we may assign the \emph{3p}-decay energy of the $^{31}$K g.s.\ as 2.15(15)+2.45$^{+0.05} \!\!\!\!\!\!\!\!\!\!\!\!_{-0.10}\simeq$4.6(2)~MeV.

Similar angular $\theta_{p-^{28}S}$ projections made with the 9 and 16 MeV gates (see Fig.~\ref{fig:theta-p-S}) are less conclusive. One may see that both distributions in Fig.~\ref{fig:theta-p-S} (ii) and (iii) contain no contribution from the \emph{2p}-decay of $^{30}$Ar g.s., and therefore the 9 and 16 MeV excited states in $^{31}$K should proceed via excited states in $^{30}$Ar.

%-------------------------------------------------------------------------------
\begin{figure}
\begin{center}
\includegraphics[width=0.47\textwidth]{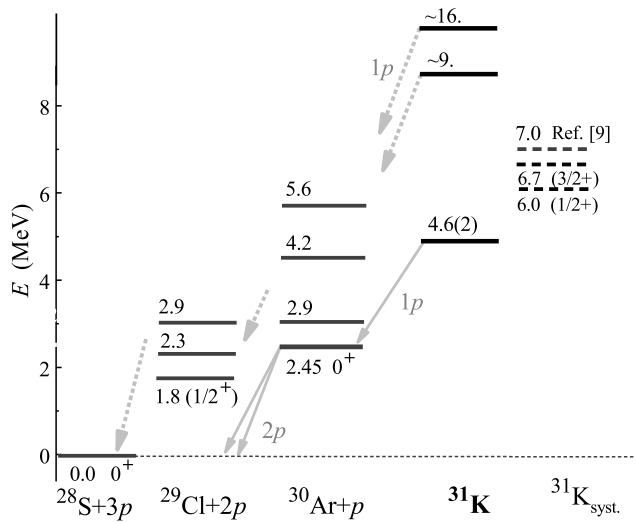}
%{aa-decay-scheme_1}
\end{center}
\caption{ Proposed decay scheme of  $^{31}$K levels with a tentatively-assigned \emph{1p}-decay channel through the known $^{30}$Ar and $^{29}$Cl states \cite{Xu:2018},
whose energy is given relative to the $3p$, $2p$ and $1p$ thresholds, respectively.
On the right-hand side, the  energies of the $^{31}$K g.s.\ predicted by the improved Kelson-Garvey mass relations  \cite{Tian:2013}, and the estimates for the tentative (3/2$^+$) and (1/2$^+$) states based on mirror energy differences  \cite{Fortune:2018} are shown by the dashed lines. }
 \label{fig:decay-scheme}
\end{figure}
%-------------------------------------------------------------------------------

The result of the data analysis, the assigned levels of $^{31}$K and their decay scheme, is shown in Fig.~\ref{fig:decay-scheme}.
The derived information may be improved in following experiments with higher statistics and increased resolution. The mass of the $^{31}$K g.s.\ may be derived by using the masses of $^{28}$S+\emph{3p} and the $Q_{3p}$ value, which then may be compared with available theoretical predictions.
The  energy of the $^{31}$K g.s.\ has been predicted by the systematics proposed for the mass differences of mirror nuclei (the improved Kelson-Garvey  mass relations \cite{Tian:2013}), which is shown in Fig.~\ref{fig:decay-scheme} on the right-hand side. One sees quite a large disagreement with the experimental value. Such a difference may be explained by the effect of Thomas-Ehrmann shift \cite{Thomas:1952,Ehrman:1951} which is often observed in \emph{1p}-unbound nuclei. Alternatively, as the $^{31}$K g.s.\  decays via the long-lived $^{30}$Ar g.s., we may use the empirical $S_p$ systematics  of \emph{1p}-emitting  states in light nuclei based on parametrization of experimental mirror energy differences (MED) \cite{Fortune:2018}. The definition is MED=$S_n$(neutron-rich nucleus)$-S_p$(its proton-rich mirror), and MED=(\emph{Z/A}$^{1/3}$)MED', where the MED' value does not depend on the nuclear charge \emph{ Z} and mass \emph{A} \cite{Fortune:2018}. This parametrization can be  scaled  to the presumably $d_{3/2}$ g.s.\ of $^{31}$K by using the known thresholds of the \emph{A}-2 mirror pair $^{29}$Cl$^*$(3/2$^{+}$)--$^{29}$Mg$_{g.s.}$(3/2$^{+}$), which results in not much better agreement with the data (see Fig.~\ref{fig:decay-scheme} on right-hand side).
The similar estimate can be done by using  the \emph{A}-2 mirror states  $^{29}$Cl$_{g.s.}$(1/2$^{+}$)--$^{29}$Mg$^*$(1/2$^{+}$) where excitation energy of the experimentally identified $^{29}$Mg$^*$(1/2$^{+}$) is 55 keV according to Ref.~\cite{Baumann:1989}. Then the evaluated $S_{3p}$ value is of -6.0 MeV which  still disagrees with the data.
Such a difference in  the observed and predicted energies of the $^{31}$K g.s.\ requires further investigation, for example the influence of three-nucleon forces may be studied like in Ref.~\cite{Holt:2013}.
%The most-matching prediction has been done
%by using measured binding energies of  the  isospin-analog  neutron-rich  nuclei  and  Coulomb  energy  shifts
%deduced  from  a  parametrization  of  measured Coulomb displacement energies, where the $^{31}$K g.s.\ was predicted to have
% $S_p$=-4.5(1) MeV \cite{Cole:1998}.
%

The width of the $^{31}$K g.s.\ derived by the fit in Fig.\ \ref{fig:theta-p-S} (i) provides only the upper-limit value $\Gamma_{g.s.}<$ 400~keV, as it reflects the experimental resolution. For comparison, the upper-limit Wigner estimate for a single-particle 1$d_{3/2}$-shell width of $^{31}$K g.s.\ is about 30 ~keV only.
The widths of the excited 9 and 16 MeV states were estimated from the $\rho_3$ distribution in Fig.\ \ref{fig:31K-rho3} giving the values of 1 and 2 MeV, respectively.
One should also note, that the spectrum of $^{31}$Mg, which is the mirror nucleus of $^{31}$K, displays a number of low-energy levels assigned to two rotational bands and to a spherical configuration \cite{Nishibata:2017}, which provides an evidence of shape coexistence in this nuclear system. Most of these states de-excite by $\gamma$-ray emission with half-lives in the nanosecond range. As all $^{31}$K states are unbound, the isospin-symmetrical rotational bands are unlikely to be excited.

%-------------------------------------------------------------------------------
\begin{figure}
\begin{center}
\includegraphics[width=0.48\textwidth]{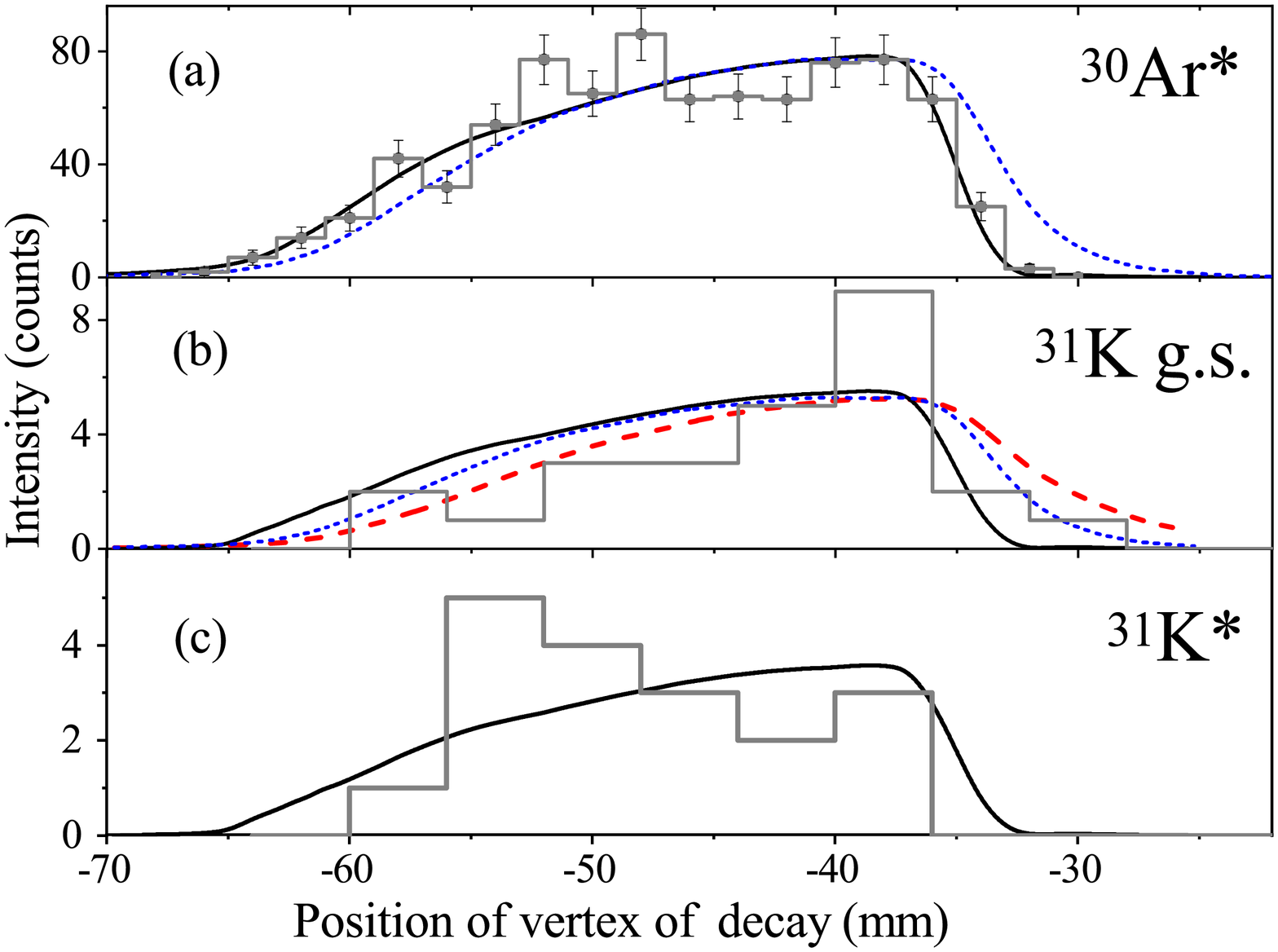}
\end{center}
\caption{ (a) Profile of the $^{30}$Ar$^*\rightarrow^{28}$S+$p$+$p$ decay vertices along the beam direction with respect to the  microstrip detector (histogram with statistical uncertainties) closest to the reaction target. The data correspond to short-lived excited states in $^{30}$Ar$^*$ \cite{Mukha:2015}. (b,c) Profiles of the $^{31}$K$\rightarrow^{28}$S+$p$+$p$+$p$ decay vertices measured under the same conditions as those shown in panel (a). In  panel (b), the data with the lowest $\rho_3$ angles around the peak (i) in Fig.~\ref{fig:31K-rho3} are selected, where the ground state of $^{31}$K is assigned. (c) The data are gated by the larger angles $\rho_3$ around the peak (ii) in Fig.~\ref{fig:31K-rho3}, which corresponds to a short-lived excited state in $^{31}$K. Solid, dotted and dashed curves show the Monte Carlo simulations of the detector
response for the  \emph{2p}-decays of $^{30}$Ar and  \emph{3p}-decays of $^{31}$K with half-life $T_{1/2}$ of 0, 5 and 10~ps, respectively.}
 \label{fig:ar30-K31-z}
\end{figure}
%-------------------------------------------------------------------------------

Nevertheless, we have evaluated the half-life values of the observed $^{31}$K states in the picosecond range by measuring distributions of their decay vertexes.  Figure \ref{fig:ar30-K31-z}~(a) shows the
profile of \emph{2p}-decay vertexes of the $^{30}$Ar$^*$ short-lived excited states first published in \cite{Mukha:2015} by using  the measured $^{28}$S+$p$+$p$ trajectories.
This profile serves as a reference in the evaluation of the  \emph{3p}-decay vertexes from the $^{31}$K ``ground'' and ``first excited'' states  which are shown in Fig.\ \ref{fig:ar30-K31-z} (b) and (c), respectively. The two latter profiles are derived from  the measured $^{28}$S+$p$+$p$+$p$ events by applying the selection gates  around the peaks (i) and (ii) in the  $\rho_3$-spectrum of $^{31}$K shown in Fig.~\ref{fig:31K-rho3}.
The Monte Carlo simulations \cite{Agostinelli:2003,Mukha:2007} of the reference case of $^{30}$Ar short-lived excited states are shown in Fig.~\ref{fig:ar30-K31-z}(a). They assume $T_{1/2}\simeq$0 ps for the $^{30}$Ar  states and take into account the experimental angular uncertainties in tracking the fragments and reconstructing the vertex coordinates. The simulations
reproduce the data quantitatively. The half-life uncertainty is illustrated by the $T_{1/2}$=5~ps simulation which fails fitting the data.
%with more than 90\% probability,
%according to the standard statistical Kolmogorov test [15].
The asymmetry of the rising and falling slopes of the
vertices is due to multiple scattering of the fragments in the thick target.
%This vertex profile serves as the reference for
%estimating the half-life of the $^{30}$Ar g.s.
%
Similar Monte Carlo simulations with $T_{1/2}$ of 0, 5 and 10~ps of the $^{31}$K states are compared with the corresponding
data  in Fig.\ref{fig:ar30-K31-z}(b).
One may see that the $T_{1/2}$=5~ps simulation is the best fit for the $^{31}$K g.s.\ data. However, production of few events of $^{31}$K  inside the 27-mm thick secondary target result in the $T_{1/2}$ uncertainties of 10~ps.   Thus we conclude that the  half-life value of the $^{31}$K g.s.\  is shorter than 10~ps, which is our upper-limit estimate. For the $^{31}$K excited-state, the best fit of its vertex profile is shown in Fig.\ref{fig:ar30-K31-z}(c) giving $T_{1/2}$=0~ps.
Though we found no indication on long-lived states in $^{31}$K, a dedicated experiment with improved resolution and larger statistics inspecting possible shape coexistence in $^{31}$K may provide a strict test of isospin-symmetry conservation/violation of such exotic nuclear systems.

In conclusion, the first spectroscopy of the previously-unknown isotope $^{31}$K, located four atomic mass units beyond the proton drip line, has revealed states whose widths are much smaller than the values of 3--5 MeV  which are mandatory for the formation of a nuclear state
\cite{Grigorenko:2018}. Therefore the half-lives of the observed $^{31}$K states are much longer than those predicted at the limits of existence of nuclear structure, and one can conclude that a transition region to chaotic nuclear systems is not reached yet. Looking to the future, similar charge-exchange reactions with more exotic beams like $^{48}$Ni or $^{67}$Kr prospect investigations of nuclear systems located by seven mass units beyond the proton drip line, where the basic mean-field concept and Pauli principle may be ultimately tested.  Last but not least, the mass of the $^{31}$K g.s.\ can be derived from the measured $S_{3p}$ value, which is the most challenging test of predictions of nuclear mass models.

%\begin{acknowledgments}
This work was supported in part by the Helmholtz International Center for FAIR (HIC for FAIR); the Helmholtz Association (Grant No.\ IK-RU-002); the Russian Science Foundation (Grant No.\ 17-12-01367); the Polish National Science Center (Contract No.\ UMO-2015/17/B/ST2/00581); the Polish Ministry of Science and Higher Education (Grant No.\ 0079/DIA/2014/43, Grant Diamentowy); the Helmholtz-CAS Joint Research Group (Grant No. HCJRG-108); the Ministry of Education \& Science, Spain (Contract No.\ FPA2016-77689-C2-1-R); the Hessian Ministry for Science and Art (HMWK)through the LOEWE funding scheme; the Ministry of Education, Youth and Sports, Czech Republic (Projects  LTT17003 and LM2015049); the Justus-Liebig-Universit\"at Giessen (JLU) and the GSI under the JLU-GSI strategic Helmholtz partnership agreement; DGAPA-PAPIIT UNAM IA103218. This work was carried out in the framework of the Super-FRS Experiment Collaboration. This article is a part of the Ph.D.\ thesis of D.~Kostyleva.
%\end{acknowledgments}

%###############################################################################

\bibliographystyle{apsrev4-1}
\bibliography{/u/dkostyl/PRL_paper_K/references/all}
%\bibliography{d:/latex/all}

%###############################################################################

\end{document}